# Deriving Analytical Solutions Using Symbolic Matrix Structural Analysis: Part 1 – Continuous Beams

Vagelis Plevris[1*] and Afaq Ahmad[2]

[1] Department of Civil and Environmental Engineering, Qatar University
P.O. Box: 2713, Doha, Qatar
e-mail: vplevris@qu.edu.qa

[2] Department of Built Environment, Oslo Metropolitan University
P.O. Box 4, St. Olavs Plass, NO-0130, Oslo, Norway
e-mail: afahm2637@oslomet.no

*Corresponding author

**Abstract.** This study investigates the use of symbolic computation in Matrix Structural Analysis (MSA) for continuous beams, leveraging the MATLAB Symbolic Math Toolbox. By employing symbolic MSA, analytical expressions for displacements, support reactions, and internal forces are derived, offering deeper insights into structural behavior. This approach facilitates efficient and scalable sensitivity analysis, where partial derivatives of outputs concerning input parameters can be directly computed, enhancing design exploration. The development includes an open-source MATLAB program, hosted on GitHub, enabling symbolic analysis of continuous beams subjected to point and uniform loads. This approach is invaluable for both engineering practice and pedagogy, enriching the understanding of structural mechanics and aiding in education by illustrating clear parameter relationships. The program supports deriving influence lines and identifying maximum response values.

## 1 Introduction to Symbolic Matrix Structural Analysis for Structural Engineering

The Finite Element Method (FEM), which recently marked its 80[th] anniversary since its inception [1], has been a foundational tool in structural analysis, offering a robust numerical framework for solving complex engineering problems [2]. Traditional FEM approaches are predominantly numerical, relying on computations to approximate structural behavior under specified loads and boundary conditions. While effective for producing detailed, problem-specific results, these methods have limitations, particularly in flexibility and generality. Numerical FEM solutions are generally applicable to a specific set of inputs, such as material properties, geometry, and external loads [3]. Modifications to these inputs necessitate a complete re-computation of the system, which can be both time-consuming and computationally intensive, especially for large-scale structures. Additionally, numerical results often obscure the relationships between key parameters, making it difficult to develop a deeper understanding of structural behavior.

FEM applied to linear-type structures, such as beams, trusses, and frames, is commonly referred to as Matrix Structural Analysis (MSA). Both methods share the same underlying principles and origins. In MSA, analytical expressions for the stiffness matrix of each element can often be derived directly, bypassing the need for numerical integration techniques such as Gaussian quadrature. This feature enhances the process by making the derivation simpler and more insightful. One of the primary drawbacks of traditional numerical FEM and MSA is their reliance on predefined boundary conditions and loading scenarios. For every new condition, the entire model must be regenerated and recalculated, leading to a repetitive and time-consuming workflow. This limitation hinders the method's adaptability and restricts its application in tasks requiring real-time analysis or optimization [4], where multiple configurations need to be evaluated quickly and efficiently [5].

Symbolic computation offers a promising alternative to these challenges. Unlike numerical computation, which yields specific numerical values, symbolic computation allows for the manipulation and solution of mathematical expressions in their exact, algebraic form. This approach opens new avenues for deriving analytical solutions and gaining deeper insights into structural behavior.

MATLAB is extensively utilized in structural engineering due to its robust numerical algorithms, which enable the resolution of a variety of engineering challenges, including matrix analysis of structures [6-8], structural dynamics [9-11], structural optimization [12-14], and more [15]. While MATLAB is primarily known for its numerical computing capabilities, it also includes features for symbolic computation. The MATLAB Symbolic Math Toolbox [16] significantly enhances MATLAB's functionality, allowing it to handle symbolic computation [17, 18]. The toolbox supports a wide range of symbolic operations, including algebraic simplifications, differentiation, integration, equation solving, and matrix manipulation [19]. It is particularly valuable in engineering, physics, and mathematics, where it facilitates the analysis of complex systems by providing exact, parameterized solutions that can be easily manipulated and interpreted. For researchers and educators, the Symbolic Math Toolbox is an indispensable resource, as it enhances the ability to explore theoretical concepts, derive closed-form solutions, and present results in a more intuitive and generalizable manner. Its integration with the program's numerical environment also allows for seamless transitions between symbolic and numerical analysis, offering a comprehensive platform for tackling both theoretical and practical problems.

While other software and programming languages like Mathematica [20], Maple [21], and SymPy (Python) [22] also offer symbolic computation capabilities, this study focuses on the MATLAB Symbolic Math Toolbox [16, 23], which was used to develop the symbolic analysis open-source code of this work. However, similar programs can be developed using these alternative platforms, following the same underlying principles. In structural analysis, the toolbox facilitates the expression of MSA solutions in symbolic form, where the system's response is represented by algebraic expressions involving key parameters related to material or section properties. This method is beneficial for analyzing structural elements such as beams, trusses, and frames, where the stiffness matrix for each element can be derived symbolically, eliminating the need for complex numerical integration. By combining the symbolic stiffness matrices of individual elements and expressing the force vector symbolically, fully symbolic solutions for displacements, element internal forces, or support reactions can be obtained. This enables flexible manipulation of solutions to accommodate various loading conditions, boundary constraints, or material properties without the need for repeated recalculations.

## 1.1 Literature Review

There have been relatively few attempts in the literature to address stiffness matrices symbolically. Korncoff and Fenves [24] made an early effort to develop a symbolic processor aimed at assisting in the generation of stiffness matrices for finite elements, despite the limited computational resources available at the time. Their results, however, highlighted several promising avenues for future research, both within the specific domain of finite element analysis (FEA) and in the broader application of symbolic processing techniques. Leff and Yun [25] presented a system for generating global stiffness matrices where elements are expressed as functions of shape parameters. Their approach builds on STRUDL syntax, but with a significant difference: instead of fixed values, joint coordinates, material properties, and forces are entered as parameterized expressions. The resulting stiffness matrix retains these parameter dependencies, providing flexibility in the analysis. This method allows engineers to explore how changes in structural parameters affect the stiffness matrix, offering a more versatile and adaptive framework for FEA, particularly beneficial for sensitivity studies and parametric design exploration.

Nagabhushanam et al. [26] developed a specialized symbolic manipulation package in FORTRAN for generating elemental matrices in FEA. The package allows users to perform symbolic manipulations through simple, user-friendly commands. With a modular structure and minimal memory requirements, it efficiently handles large-order matrices, even on personal computers. The package supports various element geometries and shape functions, including isoparametric elements, and offers a multilevel operator facility to perform several manipulations with a single command. This compact tool has been successfully applied to generate elemental stiffness, flexibility, and nodal force matrices, demonstrating its utility in symbolic FEA. Tummarakota and Lieh [27] addressed the need for efficient algorithms to model and predict the behavior of multibody structural systems in applications such as aerospace, robotics, and automotive engineering. They utilized a computer-aided symbolic method to formulate equations of motion based on Lagrange's method, offering greater physical insight compared to traditional numerical approaches. The generated equations are automatically converted into FORTRAN code, enabling simulations and control synthesis. The study demonstrated the effectiveness of this approach through two examples: a slider-crank mechanism and an aircraft model, which are solved using the Runge-Kutta-Fehlberg method for numerical integration.

Eriksson and Pacoste [28] discuss the use of symbolic software in developing finite element procedures, particularly for complex problems involving higher-order instabilities requiring precise formulations. The authors highlight that symbolic tools enhance the efficiency and clarity of procedural development, enabling effective comparisons between various element assumptions. The research includes beam formulations for plane and space models, allowing analytical verification of equivalence between displacement and co-rotational contexts. Symbolic derivation also simplifies handling finite space rotations and supports the systematic derivation of local displacements from global variables. Amberg et al. [29] describe the development of a toolbox in Maple for generating finite element codes from symbolic mathematical specifications, facilitating 1D, 2D, and 3D simulations. This toolbox has significantly accelerated research in areas such as thermocapillary convection, welding, and crystal growth by reducing the time from conceptualization to a functioning simulation to mere hours. The approach promotes flexibility, transparency, and thorough documentation, enabling researchers to easily modify models and focus on physical insights and numerical properties while minimizing errors and debugging.

Pavlovic [30] discussed the use of symbolic computation in structural engineering, highlighting its emergence as a powerful tool alongside traditional numerical methods. He reviewed past applications where symbolic computations have been underutilized and emphasized the potential for significant advancements in areas of classical structural analysis. The author argues that symbolic computation invigorates classical analytical techniques, offering new opportunities to address complex structural problems. He proposes a more balanced approach that integrates both symbolic and numerical methods, demonstrating their complementary strengths in solving structural mechanics problems efficiently and accurately.

Skrinar and Pliberšek [31] derived a symbolic stiffness matrix and load vector for a slender beam with multiple transverse cracks under uniform loading. Using the principle of virtual work, they provided closed-form analytical expressions that facilitate fast and straightforward evaluations. This approach, which excludes shape functions, makes the impact of crack depth and location on flexural deformation clearer, aiding in crack identification. The developed matrix is ideal for modeling flexural cracks in beams and columns, which is relevant in earthquake engineering per European design code EC8. Roque [32] explored the symbolic and numerical analysis of bending plates using MATLAB for symbolic manipulation of expressions. The author emphasized the importance of integrating numerical and symbolic approaches in problem-solving, highlighting MATLAB's versatility in seamlessly combining these computational methods.

## 1.2 Advantages of Symbolic Expressions and Solutions in MSA

The symbolic representation of structural analysis problems provides engineers and researchers with deeper insight into structural behavior. By maintaining algebraic relationships between parameters, symbolic MSA enables the exploration of how changes in material properties, geometry, or loading affect the overall response of the structure. Symbolic solutions are particularly useful in scenarios where flexibility and adaptability are crucial, as they provide parameterized solutions not confined to specific input values. Furthermore, when symbolic expressions are available, engineers can easily calculate partial derivatives with respect to various input parameters, facilitating sensitivity analyses and design optimization in a straightforward and efficient manner.

An additional advantage of symbolic MSA lies in its educational value. For teaching structural engineering concepts, symbolic solutions can help students and professionals better understand the fundamental principles of mechanics and structural behavior. The symbolic form explicitly reveals how physical parameters such as the modulus of elasticity or the moment of inertia of the beam's cross-section influence displacements, stresses, and strains. This level of transparency fosters a deeper conceptual understanding compared to purely numerical approaches, which often provide results without revealing the underlying mechanics. Consequently, symbolic MSA serves as both a powerful research tool and an effective educational resource for conveying complex ideas in a clear and accessible way.

### 1.3 Key Features and Novelty of this Study

This study presents an innovative, open-source MATLAB program specifically designed for symbolic MSA of continuous beams subjected to point and uniform loads. For the first time, this program enables the accurate and efficient derivation of analytical solutions for any continuous beam configuration, providing engineers with symbolic expressions for displacements, support reactions, and internal forces. These capabilities surpass traditional methods by offering solutions that can be adapted to any beam setup with ease.

Users can define problems to obtain analytical solutions for specific output quantities at any point $x$ along the beam or determine the maximum values of these quantities. The program also facilitates the generation of influence lines through symbolic expressions and supports sensitivity analysis by calculating partial derivatives of outputs with respect to input parameters. The main features of the program include:

- Closed-form solutions for any output quantities (e.g., support reactions) across various continuous beam configurations.

- Solutions expressed as functions of $x$ for specific points on the beam (e.g., bending moment $M(x)$) and determination of maximum values along the beam (e.g., $M_{max}$).

- Analytical expressions for influence lines for specified outputs (e.g., support reactions influenced by a unit load moving along the beam).

- Sensitivity analysis for any output quantity relative to input parameters, utilizing MATLAB's built-in differentiation capabilities.

The source code is freely available at GitHub (https://github.com/vplevris/SymbolicMSA-Beams), accompanied by all examples discussed in the study. This work was implemented using MATLAB R2024b and its symbolic toolbox, and it is anticipated to be compatible with earlier versions. Users are encouraged to download the code, experiment, and create their own analytical solutions. The implementation is user-friendly, requiring minimal setup to execute custom analyses.

## 2  Continuous Beam Stiffness Matrix

The stiffness matrix plays a crucial role in MSA as it defines the relationship between the applied forces and the resulting displacements within a structural system. Typically, in MSA/FEM stiffness matrices are computed numerically for each individual element and subsequently assembled into a global system of equations representing the entire structure. Numerical integration is usually needed, for example for the case of 2D plane stress elements or more complex elements, which is regularly done using Gauss quadrature. However, in MSA, for linear-type elements, it is usually possible to derive an exact symbolic expression for the stiffness matrix of an element. One such case is the 2D Euler–Bernoulli beam with 3 degrees of freedom (DOFs) per node [15] which is commonly used in the linear static analysis of plane beams and frames. In this study, we use a beam element with 2 DOFs per node, one translational and one rotational, resulting in 4 DOFs for the element, as shown in Figure 1. The axial degree of freedom is omitted because the specific element is primarily designed for the analysis of continuous beams, where axial effects are typically negligible.

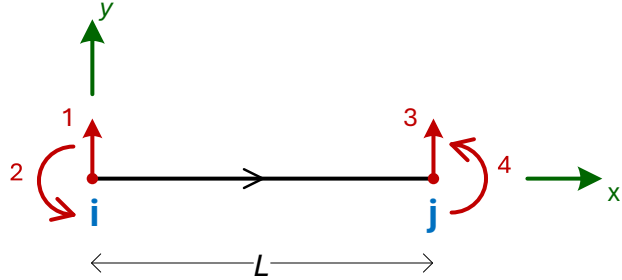

Figure 1. 2D Euler–Bernoulli beam element with 4 DOFs.

In this context, calculating the element stiffness matrix is straightforward. The symbolic 4×4 stiffness matrix for the Euler-Bernoulli beam element, assuming the axial degree of freedom is neglected, corresponding to the DOFs shown in Figure 1, is given by Eq. (1).

$$[\hat{k}_{b2}] = \frac{EI}{L^3} \cdot \begin{bmatrix} 12 & 6L & -12 & 6L \\ 6L & 4L^2 & -6L & 2L^2 \\ -12 & -6L & 12 & -6L \\ 6L & 2L^2 & -6L & 4L^2 \end{bmatrix} \tag{1}$$

In this expression, $E$ denotes Young's modulus of the material, $I$ represents the moment of inertia of the beam's cross-section, and $L$ is the length of the beam element. For continuous beam analysis, where all beam elements are essentially horizontal and oriented from the left (Start node i) to the right (End node j), the local stiffness matrix aligns with the global stiffness matrix of the element. This eliminates the need for any element transformation, as the rotation angle of the element is zero.

## 3   Educational Benefits and Importance of Symbolic Solutions in Structural Analysis

Symbolic solutions in the context of matrix structural analysis offer unique educational advantages that go beyond the limitations of traditional numerical approaches. By retaining key parameters in symbolic form, symbolic MSA enables students and engineers to explore the fundamental relationships that govern structural behavior. This approach provides a more intuitive understanding of how structural systems respond to changes in their properties, making it an invaluable tool for teaching and reinforcing concepts in structural mechanics.

One of the primary benefits of using symbolic solutions in structural analysis is the ability to visualize how different parameters affect the overall structural response. In a traditional numerical MSA solution, students are often provided with final displacement values, reaction forces, or internal stresses without a clear connection to the underlying properties of the structure. However, in symbolic form, the solutions remain expressed in terms of parameters like $E$, $I$, and $L$, allowing students to directly observe how these variables influence the results. For example, the symbolic stiffness matrix for a continuous beam of Eq. (1), shows how stiffness increases with higher values of $E$ and $I$, or how it decreases as the span length $L$ increases. This direct relationship between structural parameters and responses provides students and engineers with a much deeper conceptual understanding of the mechanics involved.

Another illustrative example is a simply supported horizontal beam subjected to a vertical point load $P$ at midspan. A purely numerical solution would provide a displacement value at the midpoint, specific to the given load and beam properties. In contrast, a symbolic solution will express the displacement at midspan as:

$$U = \frac{PL^3}{48EI} \qquad (2)$$

In this symbolic expression, students can immediately see the direct dependence of displacement on the applied load $P$, the span length $L$, and the material and geometric properties $E$ and $I$. For instance, increasing the span length or decreasing the material stiffness results in a larger displacement. Students can also realize that doubling $L$ leads to an eightfold increase in displacement, which makes $L$ an important parameter for the displacement. The symbolic form not only reinforces the theory behind structural deflection but also provides a framework for students to experiment with different values of these parameters and predict how changes will affect the behavior of the structure.

Symbolic solutions are also beneficial in helping students grasp the concept of **superposition of effects**. For instance, in the analysis of continuous beams with multiple spans and varying loads, symbolic expressions for reactions and internal forces can be developed for each span or section of the beam. By analyzing these expressions, users can see how different loads contribute to the overall structural response. The clarity of these symbolic relationships allows for a step-by-step breakdown of how each load influences the displacement and internal forces, which can be less apparent in numerical MSA results.

Another key advantage of symbolic MSA is its ability to **enhance the teaching of design sensitivity and optimization**. In a classroom setting, students are often asked to investigate how sensitive a structure's response is to changes in material properties or geometry. Symbolic solutions provide an ideal platform for this kind of analysis. For example, by differentiating a symbolic expression for displacement with respect to Young's modulus $E$, students can see how changes in material stiffness will affect the structure's performance. This provides an intuitive understanding of the sensitivity of structural response to design changes, which is crucial for both design optimization and practical engineering applications. Sensitivity analysis will be examined in detail in Section 6.

In comparison to purely numerical solutions, symbolic expressions have the added **pedagogical benefit of clarity**. Numerical MSA solutions often provide precise but isolated results that do not offer insight into the overall behavior of the system. These results can obscure key concepts, especially for students who are new to MSA. By contrast, symbolic MSA maintains a clear and general form that highlights the relationships between different aspects of the system, such as material properties, geometry, and loading conditions. This clarity allows students to see not just the solution but also the **process** by which the solution is reached. This is critical for building a solid theoretical foundation and developing problem-solving skills.

In summary, symbolic solutions in MSA provide significant educational benefits by making the relationships between structural parameters and responses more transparent and accessible. The use of symbolic expressions helps students visualize and understand how changes in properties like material stiffness, geometry, or loading affect the structural response, reinforcing theoretical concepts that might otherwise be obscured in purely numerical solutions. This approach not only enhances learning but also equips students with the skills to perform parametric studies, sensitivity analyses, and optimization in practical engineering contexts. As a result, symbolic MSA is an invaluable tool for educators and students alike in mastering the fundamentals of structural mechanics and matrix structural analysis.

## 4    Symbolic Expressions for Intermediate Values of Internal Forces and Local Extrema

In matrix structural analysis of beams, whether conducted numerically or symbolically, the results typically include displacements (or rotations) at nodal points and element forces at the ends of each element (nodes i and j). While intermediate values for displacements or internal forces are not provided directly by the method, these quantities can be determined indirectly by applying established principles of statics.

It is quite straightforward to find the intermediate values of internal forces (shear force and bending moment) at any given point $x$ within an element, based on the known forces at the ends. We consider the general case of a beam element with uniform load $q$ on it (positive $q$ points upwards, towards the $y$ axis of the beam), as shown in Figure 4.

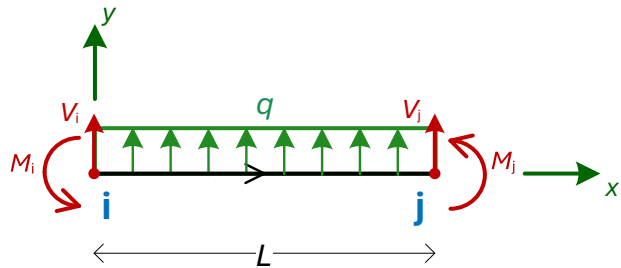

Figure 2. 2D beam element with uniform load $q$ on.

The formulas for the internal shear force and the bending moment at a given location, $x$, calculated from the left side, are:

$$V(x) = V_i + q \cdot x \qquad (3)$$

$$M(x) = -\left( M_i - V_i \cdot x - q\frac{x^2}{2} \right) = q\frac{x^2}{2} + V_i \cdot x - M_i \qquad (4)$$

Here, $x$ represents the distance along the local element $x$-axis, measured from the left node (Node i) of the element. When $x$=0, it corresponds to the starting point of the element (Node i), while $x=L$ marks the end of the element (Node j). The end moments, $M_i$ and $M_j$, are defined as positive when they act counter-clockwise, consistent with the notation in Figure 1. On the other hand, for the shear force and bending moment diagrams, the internal shear force $V(x)$ is positive when it induces clockwise rotation of the element, and the internal bending moment $M(x)$ is considered positive when it induces tension in the bottom fiber of the beam. These conventions are detailed in Figure 3(a) and Figure 3(b). In Eq. (4), the internal bending moment is calculated from the left side, necessitating a negative sign to align with this convention and ensure positivity when it induces tension in the bottom fiber, as illustrated in Figure 3(b).

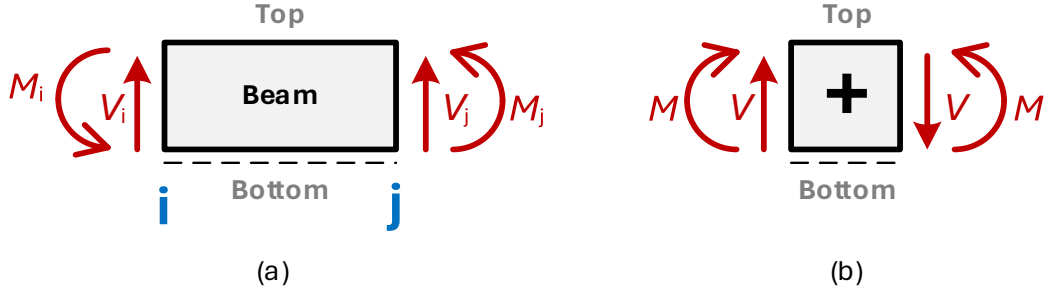

Figure 3. Positive notation used (a) For the MSA results at each end of the beam (i and j), (b) For the shear force and bending moment diagrams.

For the case of a uniform load ($q\neq0$), the extremum (maximum or minimum) bending moment is typically found at the point where the shear force equals zero, i.e., at

$$x_{extrM} = -\frac{V_i}{q} \qquad (5)$$

And according to Eq. (4) the value of the extremum bending moment will then be

$$M(x_{extrM}) = -M_i + V_i \cdot \left(-\frac{V_i}{q}\right) + w\frac{\left(-\dfrac{V_i}{q}\right)^2}{2} = -M_i - \frac{V_i^2}{2q} \qquad (6)$$

## 5   Numerical Examples

We consider five numerical examples of differing levels of complexity. In these, $EI$ is treated as a single symbolic parameter, since $E$ and $I$ consistently appear together in stiffness and other analytical expressions. The first three examples have only point loads, while the other two have uniform loads. Table 1 and Table 2 provide detailed descriptions of the five examples and the associated symbolic parameters for each.

Table 1. Details of the numerical examples 1, 2, and 3.

| Example # | Figure | Symbolic Parameters |
|---|---|---|
| 1 | 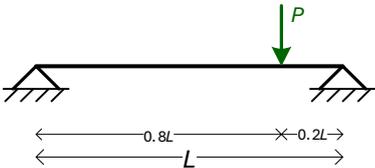 | 3<br>($EI$, $L$, and $P$) |
| 2 | 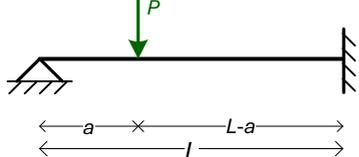 | 4<br>($EI$, $L$, $a$, and $P$) |
| 3 | 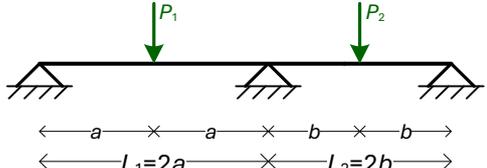 | 5<br>($EI$, $a$, $b$, $P_1$, and $P_2$) |

Table 2. Details of the numerical examples 4 and 5.

| Example # | Figure | Symbolic Parameters |
|---|---|---|
| 4 | 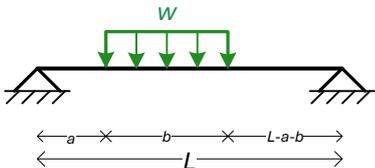 | 5<br>($EI$, $L$, $a$, $b$, and $w$) |
| 5 | 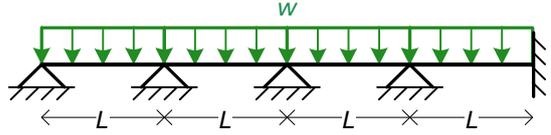 | 3<br>($EI$, $L$, and $w$) |

## 5.1 Simply Supported Beam with a Point Load (3 Symbolic Parameters)

The first numerical example is a simply supported beam with a point load $P$ applied at $x$=0.8$L$, as shown in Figure 4. The symbolic parameters are three: $EI$, $L$, and $P$.

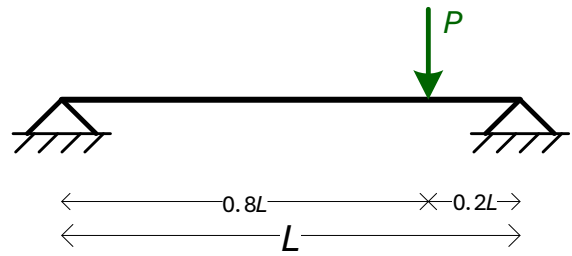

Figure 4. The beam of Example 1.

Table 3 shows the details of the model, as given in MATLAB. In every example, the number of elements (*NumElements*) is defined by the size of the *Lengths* vector and the number of nodes (*NumNodes*) is (*NumElements*-1). In this example we have 2 elements and 3 nodes. Note that a node is needed to define any point load.

Table 3. Details of the input parameters of the 1st numerical example.

| Lengths | $\begin{bmatrix} 0.8 \cdot L & 0.2 \cdot L \end{bmatrix}^T$ |
|---|---|
| Supports | $\begin{bmatrix} 1 & 0 & 1 \\ 0 & 0 & 0 \end{bmatrix}^T$ |
| PointLoads | $\begin{bmatrix} 0 & -P & 0 \\ 0 & 0 & 0 \end{bmatrix}^T$ |
| UniformLoads | $\begin{bmatrix} 0 & 0 \end{bmatrix}^T$ |

Table 4, Table 5 and Table 6 present the results of the symbolic analysis in terms of the symbolic parameters. Table 4 presents the Node displacements and rotations, while Table 5 shows the support reactions and Table 6 the element forces and moments at the start (i) and end (j) of each element.

Table 4. Example 1: Node displacements and rotations.

| Node # | y-Displacement ($D_y$) | z-Rotation ($R_z$) |
|---|---|---|
| Node 1 | 0 | $-\dfrac{4PL^2}{125EI}$ |
| Node 2 | $-\dfrac{16PL^3}{1875EI}$ | $\dfrac{4PL^2}{125EI}$ |
| Node 3 | 0 | $\dfrac{6PL^2}{125EI}$ |

Table 5. Example 1: Support reactions.

| Node # | Force $F_y$ | Moment $M_z$ |
|---|---|---|
| Node 1 | $P/5$ | - |
| Node 3 | $4P/5$ | - |

Table 6. Example 1: Element forces and bending moments.

| Element # | Start / End | Shear Force ($V_i$, $V_j$) | Moment ($M_i$, $M_j$) |
|---|---|---|---|
| 1 | Start (i) | $P/5$ | 0 |
| | End (j) | $-P/5$ | $4PL/25$ |
| 2 | Start (i) | $-4P/5$ | $-4PL/25$ |
| | End (j) | $4P/5$ | 0 |

## 5.2 Fixed-End Beam with Point Load (4 Symbolic Parameters)

The second numerical example is a fixed-end beam with a point load $P$, as shown in Figure 5. The symbolic parameters are four: $EI$, $L$, $x$, and $P$. This example will also serve to illustrate the concept of influence lines.

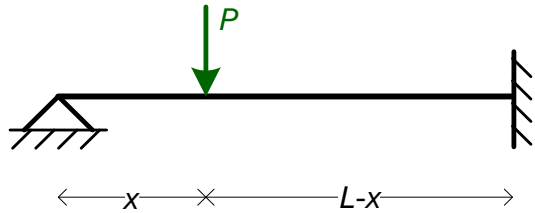

Figure 5. The beam of Example 2.

Table 7 shows the details of the model, as given in MATLAB. In this example we have 3 nodes and 2 elements.

Table 7. Details of the input parameters of the 2$^{nd}$ numerical example.

| Lengths | $\begin{bmatrix} x & L-x \end{bmatrix}^T$ |
|---|---|
| Supports | $\begin{bmatrix} 1 & 0 & 1 \\ 0 & 0 & 1 \end{bmatrix}^T$ |
| PointLoads | $\begin{bmatrix} 0 & -P & 0 \\ 0 & 0 & 0 \end{bmatrix}^T$ |
| UniformLoads | $\begin{bmatrix} 0 & 0 \end{bmatrix}^T$ |

The results of the symbolic analysis are given in Table 8, Table 9, and Table 10, for Node displacements and rotations; support reactions; and element forces and moments, respectively.

Table 8. Example 2: Node displacements and rotations.

| Node # | y-Displacement ($D_y$) | z-Rotation ($R_z$) |
|--------|------------------------|---------------------|
| Node 1 | 0 | $-\dfrac{Px(L-x)^2}{4EIL}$ |
| Node 2 | $-\dfrac{Px^2(L-x)^3(3L+x)}{12EIL^3}$ | $\dfrac{Px(L-x)^2(-L^2+2Lx+x^2)}{4EIL^3}$ |
| Node 3 | 0 | 0 |

Table 9. Example 2: Support reactions.

| Node # | Force $F_y$ | Moment $M_z$ |
|--------|-------------|---------------|
| Node 1 | $\dfrac{P(L-x)^2(2L+x)}{2L^3}$ | - |
| Node 3 | $\dfrac{Px(3L^2-x^2)}{2L^3}$ | $-\dfrac{Px(L^2-x^2)}{2L^2}$ |

Table 10. Example 2: Element forces and bending moments.

| Element # | Start / End | Shear Force ($V_i$, $V_j$) | Moment ($M_i$, $M_j$) |
|-----------|-------------|---------------------------|------------------------|
| 1 | Start (i) | $\dfrac{P(L-x)^2(2L+x)}{2L^3}$ | 0 |
| | End (j) | $-\dfrac{P(L-x)^2(2L+x)}{2L^3}$ | $\dfrac{Px(L-x)^2(2L+x)}{2L^3}$ |
| 2 | Start (i) | $-\dfrac{Px(3L^2-x^2)}{2L^3}$ | $-\dfrac{Px(L-x)^2(2L+x)}{2L^3}$ |
| | End (j) | $\dfrac{Px(3L^2-x^2)}{2L^3}$ | $-\dfrac{Px(L^2-x^2)}{2L^2}$ |

Now suppose we would like to draw the influence line of the reaction moment at Node 3. According to the results shown in Table 9, the reaction moment at Node 3 (counter-clockwise is positive) is given by

$$M_{z,3} = -\frac{Px(L^2-x^2)}{2L^2} \tag{7}$$

To draw the influence line, we assume a unit load ($P$=1) applied along the beam. For illustration purposes, we will assume that the length of the beam is also unitary ($L$=1). Then the reaction moment is given by:

$$M_{z,3} = -\frac{x(1-x^2)}{2} \tag{8}$$

The influence line corresponding to Eq. (8) is presented in Figure 6.

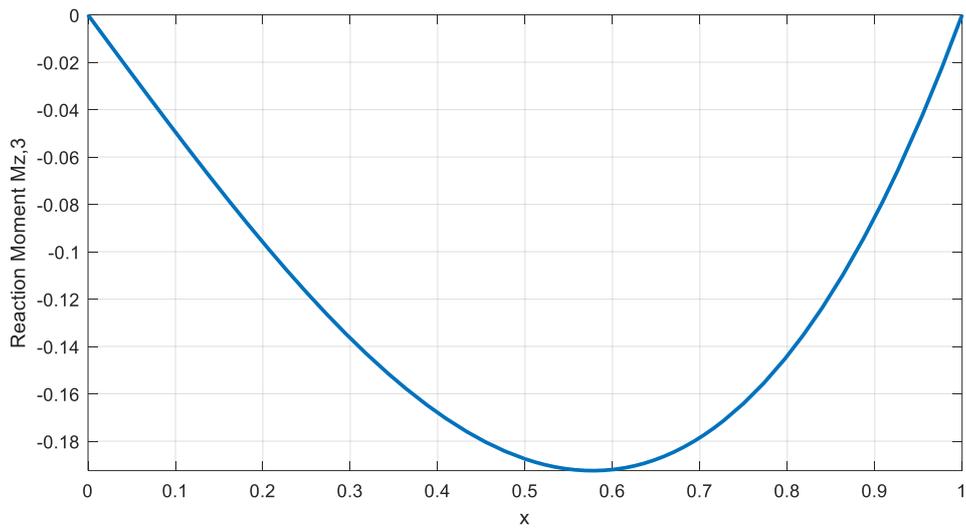

Figure 6. The influence line of the reaction moment at Node 3, for Example 3.

We observe that the reaction moment takes the value of zero when the load is at the beginning ($x$=0) or the end ($x$=$L$) of the beam, while it takes its maximum (in absolute terms) value $M_{max} = -\sqrt{3}/9 \approx -0.1925$ when the unitary load is at $x = \sqrt{3}/3 \approx 0.5774$.

### 5.3 Two-Span Beam with Two Point Loads (5 Symbolic Parameters)

The third numerical example is a two-span beam with two point loads, as shown in Figure 7. The symbolic parameters are six: $EI$, $a$, $b$, $P_1$, and $P_2$.

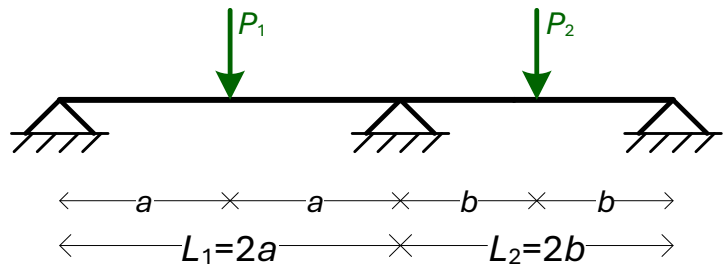

Figure 7. The two-span beam of Example 3.

Table 11 shows the details of the model, as given in MATLAB.

Table 11. Details of the input parameters of the 3rd numerical example.

| Lengths | $\begin{bmatrix} a & a & b & b \end{bmatrix}^T$ |
|---|---|
| Supports | $\begin{bmatrix} 1 & 0 & 1 & 0 & 1 \\ 0 & 0 & 0 & 0 & 0 \end{bmatrix}^T$ |
| PointLoads | $\begin{bmatrix} 0 & -P_1 & 0 & -P_2 & 0 \\ 0 & 0 & 0 & 0 & 0 \end{bmatrix}^T$ |
| UniformLoads | $\begin{bmatrix} 0 & 0 & 0 & 0 \end{bmatrix}^T$ |

The results of the symbolic analysis are reported in Table 12,
Table 13, and Table 14, for Node displacements and rotations; support reactions; and element forces and moments, respectively.

Table 12. Example 3: Node displacements and support reactions.

| Node # | y-Displacement ($D_y$) | z-Rotation ($R_z$) |
|---|---|---|
| Node 1 | 0 | $-\dfrac{a(P_1a^2 + 2P_1ab - P_2b^2)}{8EI(a+b)}$ |
| Node 2 | $-\dfrac{a^2(7P_1a^2 + 16P_1ab - 9P_2b^2)}{96EI(a+b)}$ | $\dfrac{a(P_1a^2 + P_2b^2)}{32EI(a+b)}$ |
| Node 3 | 0 | $\dfrac{ab(P_1a - P_2b)}{4EI(a+b)}$ |
| Node 4 | $-\dfrac{b^2(-9P_1a^2 + 16P_2ab + 7P_2b^2)}{96EI(a+b)}$ | $-\dfrac{b(P_1a^2 + P_2b^2)}{32EI(a+b)}$ |
| Node 5 | 0 | $\dfrac{b(-P_1a^2 + 2P_2ab + P_2b^2)}{8EI(a+b)}$ |

Table 13. Example 3: Support reactions.

| Node # | Force $F_y$ | Moment $M_z$ |
|---|---|---|
| Node 1 | $\dfrac{5P_1a^2 + 8P_1ab - 3P_2b^2}{16a(a+b)}$ | - |
| Node 3 | $\dfrac{3P_1a^2 + 3P_2b^2 + 8P_1ab + 8P_2ab}{16ab}$ | - |
| Node 5 | $\dfrac{-3P_1a^2 + 8P_2ab + 5P_2b^2}{16b(a+b)}$ | - |

Table 14. Example 3: Element forces and bending moments.

| Element # | Start / End | Shear Force ($V_i$, $V_j$) | Moment ($M_i$, $M_j$) |
|---|---|---|---|
| 1 | Start (i) | $\dfrac{5P_1a^2 + 8P_1ab - 3P_2b^2}{16a(a+b)}$ | 0 |
|  | End (j) | $-\dfrac{5P_1a^2 + 8P_1ab - 3P_2b^2}{16a(a+b)}$ | $\dfrac{5P_1a^2 + 8P_1ab - 3P_2b^2}{16(a+b)}$ |
| 2 | Start (i) | $-\dfrac{11P_1a^2 + 8P_1ab + 3P_2b^2}{16a(a+b)}$ | $-\dfrac{5P_1a^2 + 8P_1ab - 3P_2b^2}{16(a+b)}$ |
|  | End (j) | $\dfrac{11P_1a^2 + 8P_1ab + 3P_2b^2}{16a(a+b)}$ | $-\dfrac{3(P_1a^2 + P_2b^2)}{8(a+b)}$ |
| 3 | Start (i) | $\dfrac{3P_1a^2 + 8P_2ab + 11P_2b^2}{16b(a+b)}$ | $\dfrac{3(P_1a^2 + P_2b^2)}{8(a+b)}$ |
|  | End (j) | $-\dfrac{3P_1a^2 + 8P_2ab + 11P_2b^2}{16b(a+b)}$ | $\dfrac{-3P_1a^2 + 8P_2ab + 5P_2b^2}{16(a+b)}$ |
| 4 | Start (i) | $-\dfrac{-3P_1a^2 + 8P_2ab + 5P_2b^2}{16b(a+b)}$ | $\dfrac{-3P_1a^2 + 8P_2ab + 5P_2b^2}{16(a+b)}$ |
|  | End (j) | $\dfrac{-3P_1a^2 + 8P_2ab + 5P_2b^2}{16b(a+b)}$ | 0 |

## 5.4 Simply Supported Beam with Uniform Load (5 Symbolic Parameters)

The fourth numerical example is a simply supported beam with a uniform load $w$, as shown in Figure 8. The symbolic parameters are six: $EI$, $L$, $a$, $b$, and $w$.

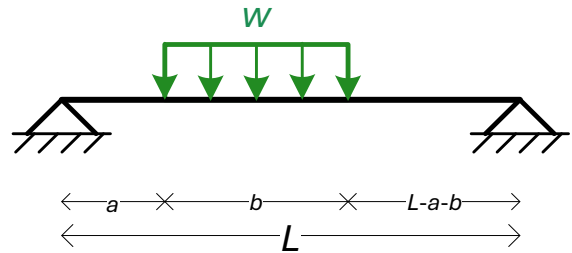

Figure 8. The beam of Example 4.

Table 15 shows the details of the model, as given in MATLAB. In this example we have 4 nodes and 3 elements.

Table 15. Details of the input parameters of the 4th numerical example.

| Lengths | $\begin{bmatrix} a & b & L-a-b \end{bmatrix}^T$ |
|---|---|
| Supports | $\begin{bmatrix} 1 & 0 & 0 & 1 \\ 0 & 0 & 0 & 0 \end{bmatrix}^T$ |
| PointLoads | $\begin{bmatrix} 0 & 0 & 0 & 0 \\ 0 & 0 & 0 & 0 \end{bmatrix}^T$ |
| UniformLoads | $\begin{bmatrix} 0 & -w & 0 \end{bmatrix}^T$ |

The results of the symbolic analysis are given in Table 16, Table 17, and Table 18, for Node displacements and rotations; support reactions; and element forces and moments, respectively.

Table 16. Example 4: Node displacements and rotations.

| Node # | y-Displacement ($D_y$) | z-Rotation ($R_z$) |
|---|---|---|
| Node 1 | 0 | $-\dfrac{bw(8L^2a+4L^2b-12La^2-12Lab-4Lb^2+4a^3+6a^2b+4ab^2+b^3)}{24EIL}$ |
| Node 2 | $-\dfrac{abw(8L^2a+4L^2b-16La^2-12Lab-4Lb^2+8a^3+8a^2b+4ab^2+b^3)}{24EIL}$ | $-\dfrac{bw(8L^2a+4L^2b-24La^2-12Lab-4Lb^2+16a^3+12a^2b+4ab^2+b^3)}{24EIL}$ |
| Node 3 | $-\dfrac{bw(2a+b)(a-L+b)(4a^2+6ab-4La+3b^2-4Lb)}{24EIL}$ | $-\dfrac{bw(2a+b)(4L^2-12La-12Lb+8a^2+14ab+7b^2)}{24EIL}$ |
| Node 4 | 0 | $-\dfrac{bw(2a+b)(-2L^2+2a^2+2ab+b^2)}{24EIL}$ |

Table 17. Example 4: Support reactions.

| Node # | Force $F_y$ | Moment $M_z$ |
|---|---|---|
| Node 1 | $-\dfrac{bw(2a-2L+b)}{2L}$ | - |
| Node 4 | $\dfrac{bw(2a+b)}{2L}$ | - |

Table 18. Example 4: Element forces and bending moments.

| Element # | Start / End | Shear Force ($V_i$, $V_j$) | Moment ($M_i$, $M_j$) |
|---|---|---|---|
| 1 | Start (i) | $-\dfrac{bw(2a-2L+b)}{2L}$ | 0 |
| | End (j) | $\dfrac{bw(2a-2L+b)}{2L}$ | $-\dfrac{abw(2a-2L+b)}{2L}$ |
| 2 | Start (i) | $-\dfrac{bw(2a-2L+b)}{2L}$ | $\dfrac{abw(2a-2L+b)}{2L}$ |
| | End (j) | $\dfrac{bw(2a+b)}{2L}$ | $-\dfrac{bw(2a+b)(a-L+b)}{2L}$ |
| 3 | Start (i) | $-\dfrac{bw(2a+b)}{2L}$ | $\dfrac{bw(2a+b)(a-L+b)}{2L}$ |
| | End (j) | $\dfrac{bw(2a+b)}{2L}$ | 0 |

The first three numerical examples did not include any uniform loads. In this fourth example, by applying the formulas of Eqs. (5) and (6) for Element 2 (with uniform load $q=-w$), we get:

$$x_{extrM} = -b \cdot \frac{2a - 2L + b}{2L} \tag{9}$$

$$M(x_{extrM}) = \frac{b^2 w(2a - 2L + b)^2}{8L^2} - \frac{abw(2a - 2L + b)}{2L} \tag{10}$$

We see that we get the exact value of the maximum bending moment along Element 2, in a symbolic form. By combining the results presented in Table 18 and Eqs. (9) and (10), we can draw the bending moment diagram of the beam in a symbolic way, including the maximum bending moment at the span 2, as shown in Figure 9.

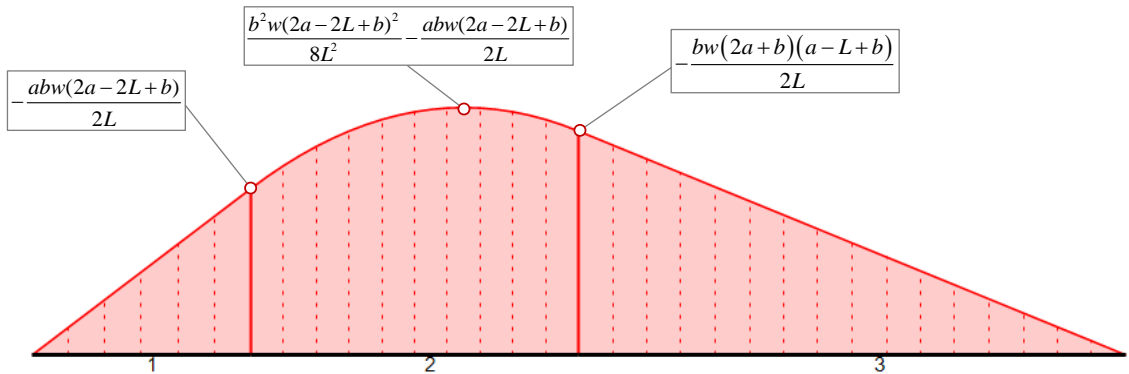

Figure 9. Symbolic bending moment diagram for Example 4.

## 5.5 Four-Span Beam with Uniform Load (3 Symbolic Parameters)

The fifth numerical example is a four-span continuous beam with a uniform load $w$, fixed at its right end, where all spans have the same length $L$, as shown in Figure 10. Although the computational model is larger and more complex than the previous ones and there are more nodes and elements, the symbolic parameters in this example are only three: $EI$, $L$, and $w$.

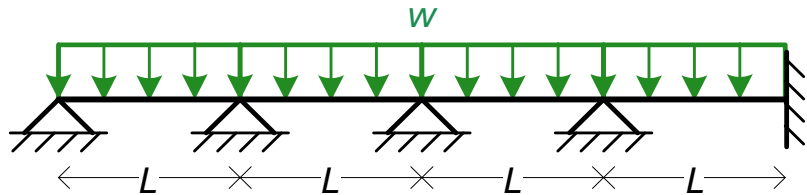

Figure 10. The four-span beam of Example 5.

Table 19 shows the details of the model, as given in MATLAB. In this example we have 5 nodes and 4 elements.

Table 19. Details of the input parameters of the 5th numerical example.

| Lengths | $\begin{bmatrix} L & L & L & L \end{bmatrix}^T$ |
|---|---|
| Supports | $\begin{bmatrix} 1 & 1 & 1 & 1 & 1 \\ 0 & 0 & 0 & 0 & 1 \end{bmatrix}^T$ |
| PointLoads | $\begin{bmatrix} 0 & 0 & 0 & 0 & 0 \\ 0 & 0 & 0 & 0 & 0 \end{bmatrix}^T$ |
| UniformLoads | $\begin{bmatrix} -w & -w & -w & -w \end{bmatrix}^T$ |

The results of the symbolic analysis are given in Table 20, Table 21, and Table 22, for Node displacements and rotations; support reactions; and element forces and moments, respectively.

Table 20. Example 5: Node displacements and rotations.

| Node # | y-Displacement ($D_y$) | z-Rotation ($R_z$) |
|---|---|---|
| Node 1 | 0 | $-\dfrac{7wL^3}{291EI}$ |
| Node 2 | 0 | $\dfrac{5wL^3}{776EI}$ |
| Node 3 | 0 | $-\dfrac{wL^3}{582EI}$ |
| Node 4 | 0 | $\dfrac{wL^3}{2328EI}$ |
| Node 5 | 0 | 0 |

Table 21. Example 5: Support reactions.

| Node # | Force $F_y$ | Moment $M_z$ |
|---|---|---|
| Node 1 | $\dfrac{153wL}{388}$ | - |
| Node 2 | $\dfrac{110wL}{97}$ | - |
| Node 3 | $\dfrac{187wL}{194}$ | - |
| Node 4 | $\dfrac{98wL}{97}$ | - |
| Node 5 | $\dfrac{193wL}{388}$ | $-\dfrac{8wL^2}{97}$ |

Table 22. Example 5: Element forces and bending moments.

| Element # | Start / End | Shear Force ($V_i$, $V_j$) | Moment ($M_i$, $M_j$) |
|---|---|---|---|
| 1 | Start (i) | $\dfrac{153wL}{388}$ | $0$ |
| | End (j) | $\dfrac{235wL}{388}$ | $-\dfrac{41wL^2}{388} \approx -0.105670wL^2$ |
| 2 | Start (i) | $\dfrac{205wL}{388}$ | $\dfrac{41wL^2}{388} \approx -0.105670wL^2$ |
| | End (j) | $\dfrac{183wL}{388}$ | $-\dfrac{15wL^2}{194} \approx -0.077320wL^2$ |
| 3 | Start (i) | $\dfrac{191wL}{388}$ | $\dfrac{15wL^2}{194} \approx -0.077320wL^2$ |
| | End (j) | $\dfrac{197wL}{388}$ | $-\dfrac{33wL^2}{388} \approx -0.085052wL^2$ |
| 4 | Start (i) | $\dfrac{195wL}{388}$ | $\dfrac{33wL^2}{388} \approx -0.085052wL^2$ |
| | End (j) | $\dfrac{193wL}{388}$ | $-\dfrac{8wL^2}{97} \approx -0.082474wL^2$ |

In this example, by applying the formulas of Eqs. (5) and (6) for all elements (as all elements have a uniform load $q=-w$ on them), we get the results presented in Table 23 for the maximum value of the internal bending moment for each span.

Table 23. Location and value of max. bending moment for each span of the beam (5[th] numerical example).

| Element # | $x_{extrM}$ | $M(x_{extrM})$ |
|---|---|---|
| 1 | $\dfrac{153L}{388}$ | $\dfrac{23409wL^2}{301088} \approx 0.077748wL^2$  * |
| 2 | $\dfrac{205L}{388}$ | $\dfrac{10209wL^2}{301088} \approx 0.033907wL^2$ |
| 3 | $\dfrac{191L}{388}$ | $\dfrac{13201wL^2}{301088} \approx 0.043844wL^2$ |
| 4 | $\dfrac{195L}{388}$ | $\dfrac{12417wL^2}{301088} \approx 0.041240wL^2$ |

* Max value at any span.

By taking the results presented in Table 22 (for element end values) and Table 23 (for span values), we can draw the bending moment diagram of the beam in a symbolic way, as shown in Figure 9.

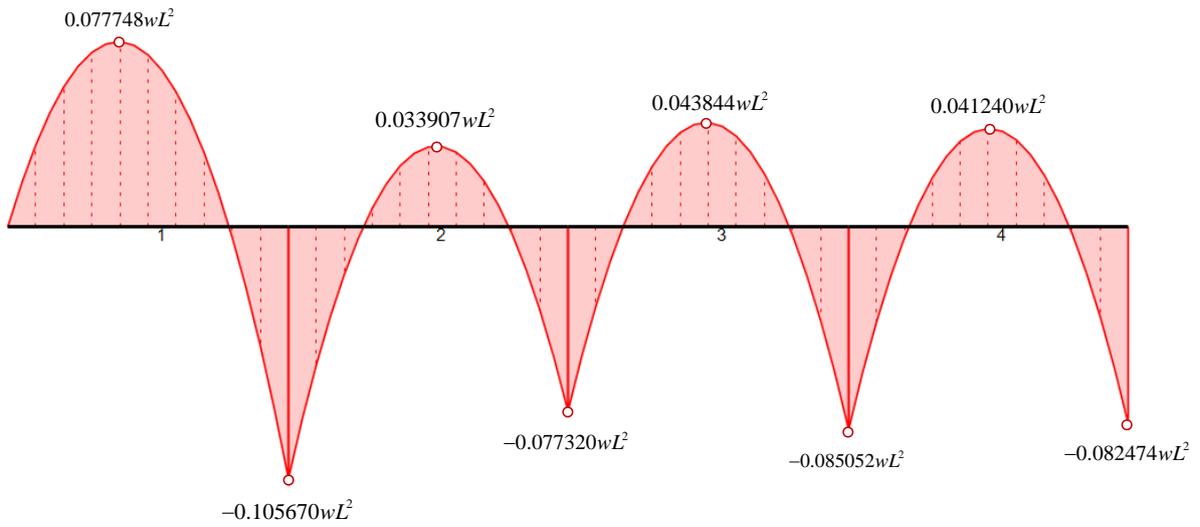

Figure 11. Symbolic bending moment diagram for Example 5.

## 6 Sensitivity Analysis

Sensitivity analysis is a powerful technique used in structural engineering to evaluate how variations in input parameters affect the system's response. It plays a critical role in understanding the robustness of a design and in optimizing performance by identifying which parameters have the most significant impact on the system's behavior. In structural analysis, engineers may want to know how changes in material properties, geometric dimensions, or applied loads affect displacements, stresses, reaction forces, or internal moments. Sensitivity analysis provides insight into these relationships, enabling more informed decision-making during the design process.

One of the major advantages of using symbolic solutions for structural engineering problems is that sensitivity analysis becomes remarkably straightforward. When the behavior of a structural system is expressed symbolically, the engineer has access to closed-form expressions that describe how key parameters such as beam length, Young's modulus, or applied loads influence the system's response. This allows for the direct calculation of **partial derivatives** of any output with respect to any input parameter. For example, with a symbolic expression for displacement, one can easily compute the sensitivity of the displacement at a specific point to changes in beam length. Similarly, the sensitivity of the internal bending moment at a given point, or a support reaction to variations in load or material properties can be computed.

For instance, consider the second numerical example shown in the previous section. The vertical displacement at $x=a$, where the load $P$ is applied, according to Table 8 is given by

$$D_{y,x} = -\frac{Px^2(L-x)^3(3L+x)}{12EIL^3} \qquad (11)$$

Using the MATLAB Symbolic Math Toolbox, we can compute the partial derivative of the displacement with respect to any parameter ($P$, $x$, $L$, $EI$), giving us a mathematical expression for how changes of a parameter affect the displacement. In this example, the partial derivative of $D_{y,x}$ with respect to $L$ can be found with the command "simplify(diff(Dyx, L))", where the MATLAB command *diff* is used for symbolic differentiation and the command *simplify* is used to simplify the expression as much as possible. The result is the following:

$$\frac{\partial D_{y,x}}{\partial L} = -\frac{Px^2(L^2-x^2)^2}{4EIL^4} \tag{12}$$

This partial derivative provides the **sensitivity** of the displacement to changes in beam length, offering valuable insight into how modifications in the design will affect the performance of the structure. The same approach can be applied to assess the sensitivity of internal forces, bending moments, or support reactions to variations in parameters like cross-sectional moment of inertia, material stiffness, or applied loads.

This ability to differentiate symbolic expressions provides a significant advantage over numerical solutions. In numerical analysis, we only obtain specific results for specific input values, making it difficult or impossible to predict how small changes in parameters will affect the outcome without rerunning the entire analysis for each scenario. Symbolic expressions, on the other hand, provide a general solution that retains the relationship between inputs and outputs, allowing for easy exploration of parameter sensitivities without additional computational cost.

Performing these calculations symbolically gives engineers an extra layer of insight, enabling them not only to see the results of an analysis but also to predict how different input parameters will influence the outputs. This ability to quickly and accurately evaluate the effects of parameter changes is especially important in the optimization phase of design, where small adjustments can significantly impact the overall performance and cost-effectiveness of a structure. This capability, which is unique to symbolic expressions, empowers engineers to predict the effects of design changes without the need for repeated numerical analysis, saving time and providing a clearer understanding of the system's behavior.

## 7 Efficiency, Scalability and Elegance of Symbolic Computations

While the symbolic implementation of MSA can offer significant advantages in terms of flexibility and insight, it also presents challenges in terms of computational efficiency and scalability, particularly for larger systems. As structural systems grow in complexity—whether through increased degrees of freedom, more elements, or complex loading conditions—the symbolic representations of stiffness matrices, force vectors, and displacement fields can become increasingly large and computationally demanding. It is important to address these challenges to ensure that symbolic MSA remains practical and efficient for engineering applications.

One of the primary issues with symbolic computations is the **exponential growth** in the size of symbolic expressions as the complexity of the problem increases. For a simple system, a symbolic stiffness matrix or equilibrium equation may provide a clear and elegant solution. However, for larger systems or those involving a large number of symbolic variables, the resulting expressions can quickly become excessively long and complex, making them difficult to manage, while losing their interpretability and usefulness for the engineer. For instance, a symbolic solution that spans multiple pages with dense expressions may offer little practical value compared to a simple numerical result. This is a critical limitation because, even though symbolic software like MATLAB will always attempt to generate a symbolic solution, the **elegance and simplicity** of the solution are paramount for engineers. A symbolic expression that is excessively long and convoluted is not only difficult to work with but also loses its value in providing insight into the system's behavior.

To manage this complexity, one effective approach is to **limit the number of symbolic variables** used in the formulation. For instance, rather than expressing every parameter of the system symbolically, it may be beneficial to identify key parameters that are most important to the analysis. By reducing the number of symbolic variables to a smaller, more manageable set, the symbolic representation remains compact and comprehensible, even for larger systems. This **hybrid symbolic-numerical approach** allows for symbolic manipulation of the most important parameters while handling the less critical parts of the system numerically. This combination of symbolic and numerical methods provides a balance between flexibility and computational efficiency, making it possible to apply symbolic MSA to larger or more complex systems without overwhelming the computational resources. As a result, the solution retains its generality and flexibility without sacrificing efficiency.

The elegance of the solution is what ultimately matters to the engineer. If the symbolic solution is compact, easy to interpret, and provides useful insights, it can be extremely valuable. On the other hand, if the solution is overly large and difficult to work with, it loses its practical utility, even if it technically provides a valid answer. Striking the right balance between symbolic and numerical methods ensures that engineers can take advantage of the benefits of symbolic computation without being overwhelmed by computational complexity.

## 8    Conclusions

This study has presented the development and application of an open-source MATLAB program capable of performing symbolic matrix structural analysis for continuous beams under point and uniform loads. This tool is freely available and provides the capability to rapidly and accurately generate analytical solutions for any beam configuration. The profound benefits of obtaining analytical solutions extend beyond practical engineering applications to enhancing engineering education, where such solutions help deepen the understanding of structural behavior.

The code goes beyond simply deriving analytical solutions for displacements, support reactions, and internal forces. It can also be used to produce influence lines for continuous beams and facilitate sensitivity analysis, which is vital for optimization and other engineering applications. The ability to compute partial derivatives of output parameters with respect to input variables enables users to assess how changes in design properties influence the overall response, making the tool valuable for both design exploration and performance evaluation.

While symbolic solutions offer significant advantages, it is essential to approach them with consideration for elegance and simplicity. The program is capable of producing large and complex symbolic expressions, but these may lack practical usefulness if they become too intricate to interpret or apply effectively. On the other hand, concise and simpler analytical solutions fulfill their intended purpose by being more comprehensible and actionable. The efficiency, scalability, and elegance of symbolic computations are crucial aspects that engineers and students must balance when employing this approach.

Through various examples, we have demonstrated the power of the code and showcased its potential to offer insightful analytical solutions. The next step in this line of research is to extend the methodology to other structural systems, such as 2D trusses and frames, to derive symbolic solutions for these structures as well. This future research direction holds promise for expanding the application of symbolic MSA and enhancing its contribution to structural analysis and education.

**Conflict of Interest**

The authors declare that the research was conducted in the absence of any commercial or financial relationships that could be construed as a potential conflict of interest.